\documentclass[intlimits,twoside,a4paper]{article}

\usepackage{amsmath,amssymb}
\usepackage{wrapfig}
\usepackage{xcolor}
\usepackage{graphicx}
\usepackage[T2A]{fontenc}
\usepackage[cp1251]{inputenc}

\usepackage{cmpj2}

\issue{2016}{19}{1}{13603}
\doinumber{10.5488/CMP.19.13603}

\title[Off-center charges]{Double layer for hard spheres with an off-center charge}

\author[W. Silvestre-Alcantara \textsl{et al}.]{W. Silvestre-Alcantara\refaddr{label1},
L.B. Bhuiyan\refaddr{label1}\thanks{E-mail: beena@beena.uprrp.edu}\,,
S. Lamperski\refaddr{label3},
M. Kaja\refaddr{label3},
D. Henderson\refaddr{label5}}
\addresses{
\addr{label1} Department of Physics and Laboratory for Theoretical Physics,
University of Puerto Rico, \\ San Juan, Puerto Rico 00936-8377, USA
\addr{label3} Department of Physical Chemistry, Adam Mickiewicz University in Pozna\'n, \\
Umultowska 89b, 61-614 Pozna\'n, Poland
\addr{label5} Department of Chemistry and Biochemistry,
Brigham Young University, Provo UT 84602-5700, USA
}

\date{Received November 9, 2015, in final form December 16, 2015}
\authorcopyright{W. Silvestre-Alcantara, L.B. Bhuiyan, S. Lamperski,
M. Kaja, D. Henderson, 2016}
\begin{document}

\maketitle

\begin{abstract}

Simulations for the density and potential profiles of the ions in the planar
electrical double layer of a model electrolyte or an ionic liquid are reported.
The ions of a real electrolyte or an ionic liquid are usually not spheres; in ionic liquids,
the cations are molecular ions.  In the past, this asymmetry has been modelled by
considering spheres that are asymmetric in size and/or valence (viz., the primitive model)
or by dimer cations that are formed by tangentially touching spheres.  In this paper we consider
spherical ions that are asymmetric in size and mimic
the asymmetrical shape through an off-center charge that is located away from the center of the
cation spheres, while  the anion charge is at the center of anion spheres. The various singlet density
and potential profiles are compared to (i) the dimer situation, that is, the constituent spheres
of the dimer cation are tangentially tethered, and (ii) the standard primitive model. The results
reveal the double layer structure to be substantially impacted especially when the cation is the
counterion. As well as being of intrinsic interest, this off-center charge model may be useful for theories that
consider spherical models and introduce the off-center charge as a perturbation.

\pacs 61.20.Qj, 82.45.Fk, 82.45.Gj, 82.45.Jn

\end{abstract}

This article is dedicated to our colleague and friend, Stefan Soko{\l}owski, in
commemoration of his 65th birthday.  DH first met and collaborated with ``Don Esteban'' in
Mexico City but had
admired his work long before that. Stefan has been our good friend and
frequent collaborator since that time.  We wish him a happy birthday and continued
good health and productivity.

\section{Introduction}

The electrical double layer (EDL) has long fascinated experimentalists and
theorists.  An EDL is formed when the charged particles in a Coulomb fluid are attracted
(or adsorbed) by a charged surface or electrode.  The charge within or on the charged
surface forms one layer and the charge of the adsorbed fluid forms the second layer that
is of opposite sign to the charge of the surface but is
equal in magnitude to the electrode charge; hence, the name double layer.   In reality,
these two layers of the EDL often consist of sublayers, so the term double layer is not quite
accurate.  Generally, the structure of the (usually metal) electrode is ignored and
the electrode is considered to be a classical metal with the (electrode) charge being
distributed uniformly on the electrode surface.  The fluid layer, on the other hand, can be
thought of as an extended diffuse layer whose net charge is equal in magnitude but opposite in sign to
the electrode charge.   The fluid layer may consist of sublayers, sometimes with alternating
charge.

The EDL has been of great practical importance in electrochemistry and analytical chemistry.
In recent years, the EDL has been found to be promising for new batteries, fuel cells, and
supercapacitors \cite{1} as well as for studies in biology \cite{2}.  The selectivity of a physiological
channel can be thought of as an EDL problem with the channel playing the role of an electrode
and the adsorbed ions playing the role of the diffuse layer \cite{3}.  The adsorption of ions by DNA
is another example of an EDL (see for example, reference~\cite{4}).

The importance of the EDL is attested to in some recent reviews  \cite{2,5,6}.
Despite this importance, the acceptance of the recent theoretical developments has
been somewhat hampered by the fact that deviations from the predictions of the classical and
often inadequate Poisson-Boltzmann (PB) theory of Gouy \cite{7}, Chapman \cite{8}, and Stern \cite{9} (GCS)
occur at high concentrations and/or high electrode charge and can be difficult to observe in
aqueous systems.  The popularity of the GCS theory stems from its intuitive simplicity and
ease of use. Thus, it might be useful in quick and qualitative analysis of experimental results
but it is unsatisfactory if one wishes to understand these results at a fundamental level.
In passing, we note that the classical theory for bulk electrolytes that parallels the GCS
theory is the more well-known Debye-H\"{u}ckel (DH) theory (see for example, reference \cite{10}).
Historically, the GCS theory predates the DH theory, however, unlike the former theory, the
latter theory is a linearized version of the bulk PB theory.

Until recently, a broad interest in the theory of the EDL has not been helped by
the fact that it is difficult to achieve experimentally the electrode charges and/or high electrolyte
concentrations with aqueous electrolytes where the deficiencies of the GCS theory are
most apparent.  By contrast, computer simulations provide a means of testing theories for such difficult
situations.  They are the gold standard against which theories can be compared since the
theory and simulation are based on the same assumed model for the inter-particle interactions.
For a given model of the electrode and ions, simulations provide exact results, apart from statistical
uncertainties, against which a theory may be compared.
Torrie and Valleau \cite{11} found in their seminal
simulations that the capacitance near the point of zero charge (pzc) and at high concentration
was greater than what was predicted by
the GCS theory.  Blum \cite{12}, and Henderson, Saavedra-Barrera and Lozada-Cassou \cite{13} showed that the
sophisticated mean spherical approximation (MSA) and hypernetted chain theory (HNC) also
exhibited this behavior.  The MSA and HNC are closely related; the MSA is a linearized version of the HNC.
A careful examination \cite{5,14} of experimental results confirms this behavior but the effect is small.
In their simulations, Torrie and Valleau also found that at high electrode charge the capacitance is smaller than the GCS
predictions.  One of us recalls a meeting at
which a prominent experimentalist ridiculed this result as being outside experimental range.  However,
the effect is still real.  Ions occupy space and their charge cannot be
crowded indefinitely. They cannot continue to form a monolayer on the electrode as the electrode
charge increases. Regrettably, the MSA and the HNC fail to predict this latter effect.
The MSA is a linear response theory and makes no prediction about the behavior of the
capacitance at large electrode charge, while the HNC proves inadequate and predicts a greater
capacitance at a large charge \cite{15}.

The modified Poisson-Boltzmann theory (MPB) \cite{16} also predicts an increasing capacitance
with an increasing concentration at small electrode charge and seems consistent with the continued
decline in the capacitance as a function of electrode charge at high electrode charge. Although
the numerical solutions of the MPB equations lack convergence at sufficiently high electrode charge,
within the range of the surface charge for which solutions exist, the capacitance curve passes
from a double hump curve at small concentrations to a single hump curve at high concentrations
consistent with simulations \cite{17,18}. In these works \cite{17,18}, the MC and MPB were applied to a
restricted primitive model (RPM) (charged hard spheres with a common diameter). The density
functional theory (DFT)\cite{19} seems to be satisfactory in a similar vein.
Here, too the DFT applied to the RPM double layer gave very good agreement with the corresponding
MC simulation results for the capacitance \cite{20}. The capacitance at small electrode
charge continues to increase with an increasing concentration but decreases with an increasing electrode
charge at large electrode charge, and shows the aforementioned double hump to single hump
transition.

Interest in the EDL at high concentrations and electrode charge
is increasing because EDLs can be formed in ionic liquids.  Ionic liquids are organic electrolytes.
Due to the fact that a solvent
is not present, ionic liquids are, in essence, room temperature molten salts.  They can exist
at high concentrations and can support higher electrode charges.  As a result, they are ideal
for the testing of modern theories of the EDL.  As Kornyshev has aptly stated \cite{20}, an ionic electrolyte
gives a new paradigm for EDL studies.  As Kornyshev observes, ionic liquid capacitance curves can show
double and single hump shapes.  There have been a variety of recent experimental studies in
which the capacitance exhibits both single and double hump shapes \cite{21,22,23,24,25} as the ion concentration
is varied.

Summarizing the situation, the differential capacitance of the EDL of an ionic liquid
at small electrode charge increases, apparently without limit, and the differential capacitance,
at large electrode charge decreases as the electrode charge increases in magnitude.
This latter effect is due to the fact that the ions cannot be adsorbed into a monolayer as the
GCS theory suggests.  These two features have the effect of causing the capacitance to pass
from a double hump shape with a minimum at or near the pzc at low concentrations to a single
hump shape at higher concentrations. By contrast, the GCS theory predicts only a minimum in
the capacitance that gradually fills in with an increasing concentration with the capacitance
gradually becoming independent of the concentration and electrode beyond some high concentration
result.

Ionic liquids, in the real world, are formed from asymmetric ions. We have considered EDLs in a
wide range of models that approximate ionic liquids  \cite{26}.  We have modelled the asymmetry
of ionic liquids by means of size asymmetry \cite{27}, through the use of dimers formed by
tangentially touching spheres \cite{27,28,29,30,31,32,33,34}, and dimers of fused spheres \cite{35}. Another method of
introducing asymmetry is to keep spherical ions but allow the charge of one species of
the ions to be off-center.  We do this by considering the charged sphere of a dimer cation
to be smaller than the diameter of the uncharged sphere of the dimer and fusing
the smaller charged sphere of the dimer cation so that it should be entirely within the larger neutral
sphere.  This results in an electrolyte consisting of spheres but with one species (the anions) having
a charge at the sphere center and the other species (the cations) having an off-center charge.
In this paper, we will report some results for the density and potential profiles of this off-center ion
model double layer.  It is to be hoped that as well as its intrinsic interest, this
model of off-center charges might lend itself to theories that use an expansion in powers
of the degree to which the charge is off-center. In the event of such theories being developed,
our simulations will be of value for testing these theories.

In addition to our work using spheres, there is another body of work that uses
simulations with fairly realistic complex ionic liquids \cite{36,37,38,39}.  The investigators using
complex ionic liquid models seek specificity whereas we seek generality. Such complex ionic
liquid models do not lend themselves to theory.  At present, only simulations are possible for these
complex models.  Both approaches have their merit.

\section{Model; simulation method}

We employ a fluid electrolyte model that consists of anions and cations that are represented by
charged hard spheres in a uniform dielectric background.  The charged spheres have a charge whose
magnitude is equal to the proton charge $e$ and have an equal diameter, $d= 4.25 \times 10^{-10}$~m.  This
model is similar to that of the restricted primitive model electrolyte but we allow the charge of
the cations to be off-center, while the charge of the anions continues to be located at the sphere center.
The electrode is considered to be a non-polarizable, hard planar wall of uniform (surface) charge density,
$\sigma$, that is located on the electrode surface.  To yield cation spheres with an off-center charge, we use
dimer cations consisting of an uncharged hard sphere and a smaller charged hard sphere, which can fuse
into each other. Thus, when the charged sphere is encapsulated entirely within the uncharged sphere,
we get a cation with an off-center charge. The anion is a charged sphere whose diameter is equal
to that of the uncharged or neutral sphere of the dimer. Specifically, if $d_{+}$, $d_{-}$, and $d_{0}$ are
the diameters of the positive, negative, and neutral spheres, respectively, then we have,
$d_{-}=d_{0}=d$ and $d_{+}=d/2$.

In the Hamiltonian, the various interaction potentials occurring in the system are
as given below. The interaction between the spheres is
\begin{equation}
 u_{ij}(r)=\left\{
\begin{array}{ll}
 \infty, & r<d_{ij}^\text{c}, \\
\frac{Z_i Z_je^2}{4\pi\epsilon_0\epsilon_r r}, & r \geqslant d_{ij}^\text{c},
\end{array}
\right.
\end{equation}
while the interaction of a sphere with the electrode is given by
\begin{equation}
 w_{s}(x)=\left\{
\begin{array}{ll}
 \infty, & x<d_{ws}^\text{c}, \\
-\frac{\sigma Z_s ex}{\epsilon _{0}\epsilon _{r}}, & x \geqslant d_{ws}^\text{c}.
\end{array}
\right.
\end{equation}
Here, $Z_{s}$ is the valency of particle species $s$,
$\epsilon_0$ and $\epsilon_r$ are the vacuum and relative permittivities, $r$ is the distance                                                                                            between the centers of two spheres, and $x$ is the perpendicular distance of a sphere from
the electrode plane.

    The quantity $d_{ij}^\text{c}$ is the distance of closest approach between two particles
of type $i$ and $j$, respectively, while $d_{ws}^\text{c}$ is the distance of closest approach
of a particle of species $s$ to the electrode. In the present study, the following three
cases may be distinguished:

\begin{itemize}
\item[(i)] for the dimer cation case (no fusion), we get as before (see for example,
reference [34]) $d_{ij}^\text{c}=(d_{i}+d_{j})/2$ and $d_{ws}^\text{c}=d_{s}/2$, $d_{s}$
being the diameter of a particle of type $s$,

\item[(ii)] for the off-centered charged cation case, we have $d_{++}^\text{c} \in [d/2,3d/2]$,
$d_{+0}^\text{c}(=d_{+-}^\text{c}) \in [3d/4,5d/4]$, and $d_{ij}^\text{c}=d$ where $ij = 00, 0-, --$. Also,
$d_{w+}^\text{c} \in [d/4,3d/4]$ and $d_{w0}^\text{c}=d_{w-}^\text{c}=d/2$, and finally

\item[(iii)] when the smaller positively charged sphere is completely fused inside the
larger neutral sphere such that the centers of the two spheres are coincident, we have
$d_{ij}^\text{c}=d$ and $d_{ws}^\text{c}=d/2$.

\end{itemize}

\begin{wrapfigure}{i}{0.5\textwidth}
\centerline{
\includegraphics[width=0.35\textwidth]{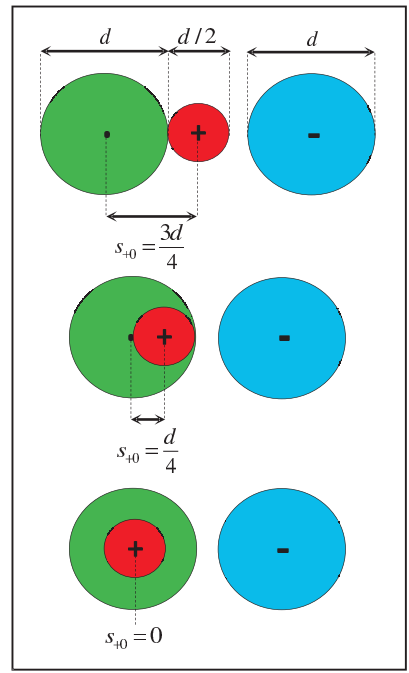}
}
\caption{(Color online) Schematic diagram of an off-center cation containing RPM.
The distance $s_{+0}$ is between the centers of the positively charged sphere and the neutral sphere.
From top to bottom: dimer model (no fusion), the off-center charge RPM,
and standard RPM, respectively. \label{fig1}}
\end{wrapfigure}
The degree to which the charge of the cations is off-center is controlled by the
separation of, or distance between, the {\it centers} of the spheres in the dimer,
$s_{+0}$.  Thus, when there is no fusion, and the spheres in the dimer are merely
tangentially touching, the separation between them is $s_{+0}=3d/4$.
We have $s_{+0}=d/4$ when the positively charged sphere is completely
inside the neutral sphere and an off-center charged cation results. This is
the case of most interest in the present work.  The $s_{+0}=0$ when
the centers of the positive and neutral spheres coincide, and the standard RPM ensues.
It ought to be mentioned, however, that although the sizes of the cation and the anion
are different, viz., $d/$2 and $d$, respectively, the fact that the positive sphere is
symmetrically embedded within the neutral sphere, makes the composite sphere an effective
hard sphere cation of the same size as the hard sphere anion, and hence the RPM.
For the diameters considered here, the off-center charge
lies between the center of the neutral sphere and a distance of $d/$4 from its center
({\it i.e.}, mid-way to the surface of the neutral sphere).  Results for an
off-center charge at a distance greater than mid-way to the surface, $d/4$, could be
obtained by using a smaller value for the diameter of the positive sphere.  However,
we expect that a value of $s_{+0}$ that is less than, or equal to, $d/4$ would be of
most interest, at least in a theory in which the degree to which the charge is off-center
was treated as a perturbation. The relation between $s_{+0}$ and the degree to which the
small charged sphere is fused into the larger uncharged sphere is shown in figure~\ref{fig1}.

The  MC simulations were performed in the canonical ($NVT$) ensemble following
the standard Metro\-polis algorithm methods. The details of the simulations
are the same as in our previous publications \cite{28,29,30,31,32,33,34,35} and will not be repeated here.
When $s_{+0}=3d/4$ and $s_{+0}=0$, that is the dimer
and RPM situations, respectively,  the simulations were checked against independent
simulations of the dimer and RPM double layers using our previous programs. These results
are useful tests of our present simulations and were found to be  consistent with our
earlier results within statistical errors.

\section{Results}

We consider a system of anions, cations, and uncharged hard spheres with a background relative
permittivity of $\epsilon_r=78.5$ at a temperature of $T=298.15$~K. The hard sphere anions and the
neutral spheres have a diameter of $d=4.25\times 10^{-10}$~m, while the
hard sphere cations are smaller (half the anion or the uncharged sphere diameters).  The
electrolyte concentration considered here is $c = 1$~mol/dm$^3$. In this paper we will restrict ourselves
to the symmetric valency cases of monovalent and divalent charged ions, viz., $Z_{+} = -Z_{-} = Z = 1$ or 2,
for the 1:1 or 2:2 cases, respectively.

In our numerical calculations, it is convenient to use reduced or dimensionless units that are
denoted by an asterisk.  However, in reporting results we will also give the equivalent physical units.
The reduced temperature is $T^*=4\pi\epsilon_0\epsilon_rdk_\text{B}T/Z^2e^2$, whose values in the present study are
$T^* = 0.595$ for the monovalent case ($Z = 1$), and $T^* = 0.149$ for the divalent case ($Z = 2$), respectively.
The reciprocal of the $T^{*}$ is the plasma coupling constant $\Gamma $ of the literature, viz., $\Gamma = 1/T^{*}$.
The bulk reduced density of the \emph{free} particles of species $s$ is defined as $\rho_s^*=\rho_{s}d^{3}$
with $\rho _{s}$ being the corresponding bulk number density. Thus, for $c = 1$~mol/dm$^3$, we have
$\rho _{+}^{*} = 0.00578$, $\rho _{-}^{*} = 0.0462$, and $\rho _{0}^{*} = 0.0462$, respectively. The reduced
surface charge density on the electrode is defined $\sigma^*=\sigma d^2/e$, while the reduced
electrode potential is $\psi^*=e\psi/k_\text{B}T$, where $\psi$ is the electrode potential in volts.

The double layer structure is described principally by the electrode--particle singlet distribution
function $g_{s}(x)=\rho _{s}^{*}(x)/\rho _{s}^{*}=\rho _{s}(x)/\rho _{s}$, where $\rho _{s}^{*}(x)$ (or
$\rho _{s}(x)$) is the local value of the corresponding quantity. In addition to reporting the values for
the $g_s(x)$ of the particles of species $s$, we will also report the values of the reduced potential,
$\psi^*(x)$. The potential profile $\psi (x)$ is a weighted integral of the singlet
distributions, viz.,
\begin{equation}
\psi(x)=\frac{e}{\epsilon_{0}\epsilon _{r}}\sum_{s}Z_{s}\rho _{s}\int_x^{\infty}\rd x'(x-x')g_{s}(x').
\end{equation}
It is a useful indicator of the overall charge distribution in the system besides being
the relevant quantity in characterizing the capacitance behaviour of the double layer.
All of the profiles are calculated as functions of the perpendicular distance, $x$ from the
electrode, which is located at $x=0$.  The reduced unit for $x$ is, of course, $x/d$.

\begin{figure}[!t]
\centerline{
\includegraphics[width=0.49\textwidth]{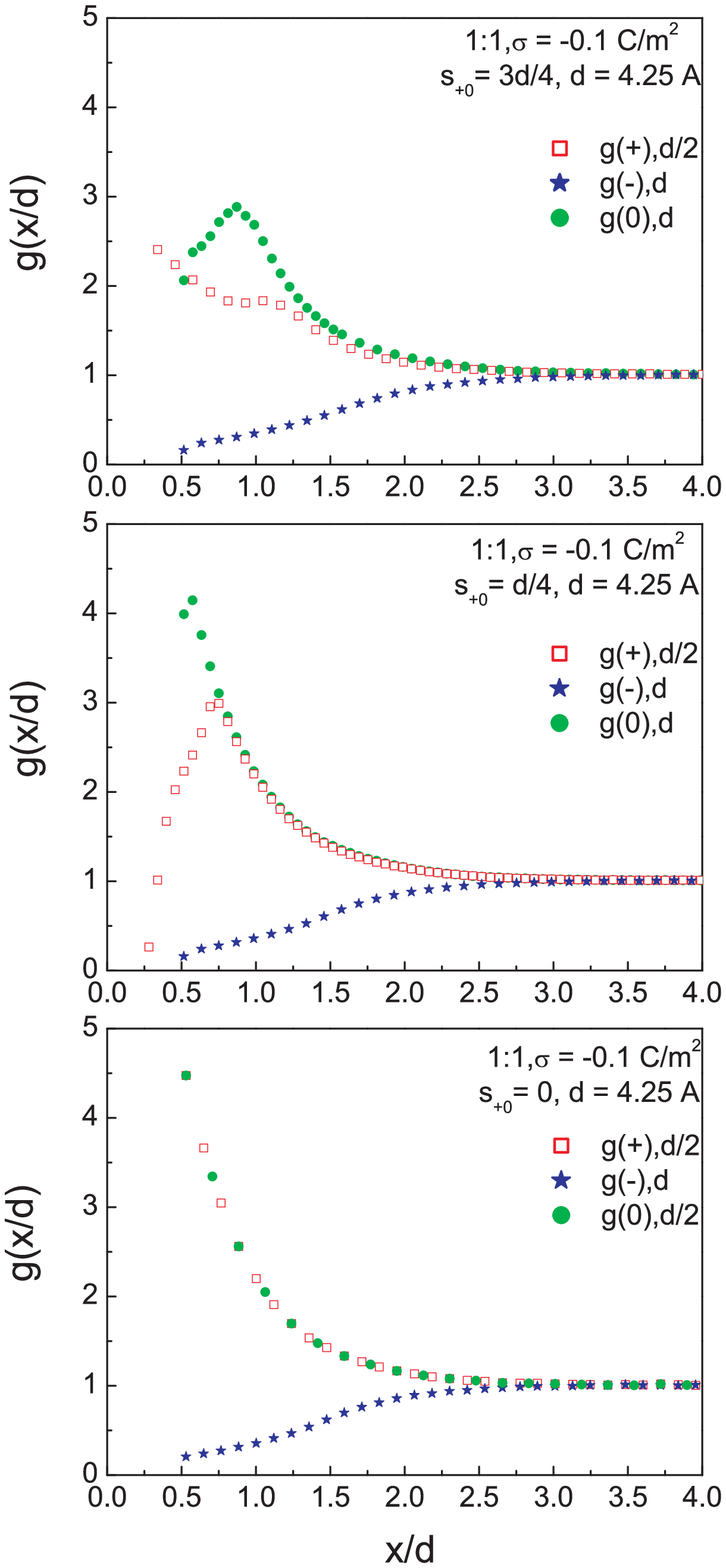}
\includegraphics[width=0.49\textwidth]{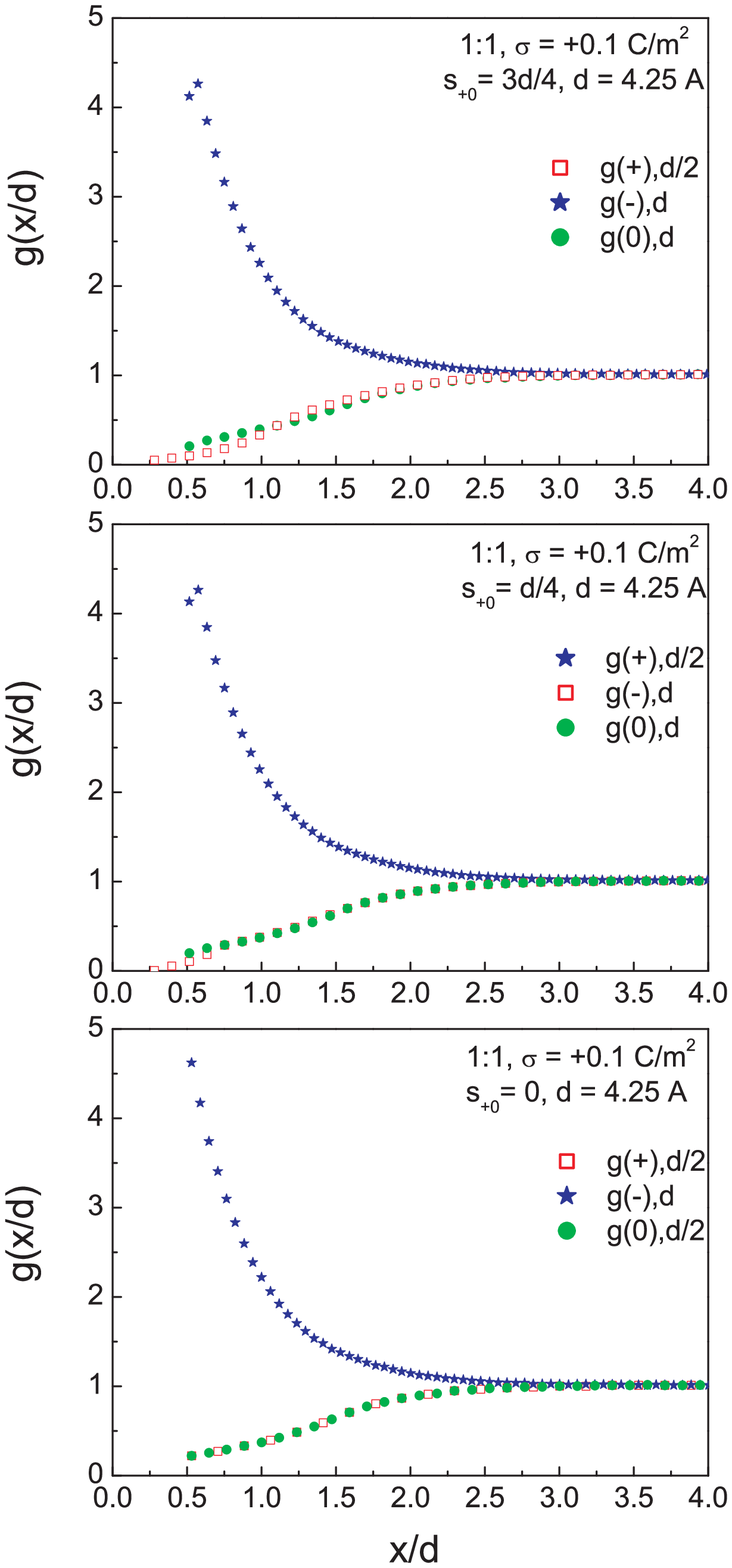}
}
\parbox[t]{0.49\textwidth}{
\caption{(Color online) Electrode--particle singlet distributions $g_{s}$ for  a 1:1 valency electrolyte system
at concentration $c = 1$~mol/dm$^{3}$, reduced electrode surface charge density $\sigma ^{*} = -0.113$
($\sigma = - 0.1$~C/m$^{2}$), and for $s_{+0}=3d/4$ (top panel), $s_{+0}=d/4$
(middle panel), and $s_{+0}=$ 0 (bottom panel). \label{fig2}}}
\parbox[t]{0.49\textwidth}{
\caption{(Color online) Electrode--particle singlet distributions $g_{s}$ for  a 1:1 valency electrolyte system
at concentration $c = 1$~mol/dm$^{3}$ and reduced electrode surface charge density $\sigma ^{*} = +0.113$
($\sigma = + 0.1$~C/m$^{2}$. Rest of the notation and legend as in figure~\ref{fig2}.
\label{fig3}}
}
\end{figure}

\begin{figure}[!t]
\centerline{
\includegraphics[width=0.5\textwidth]{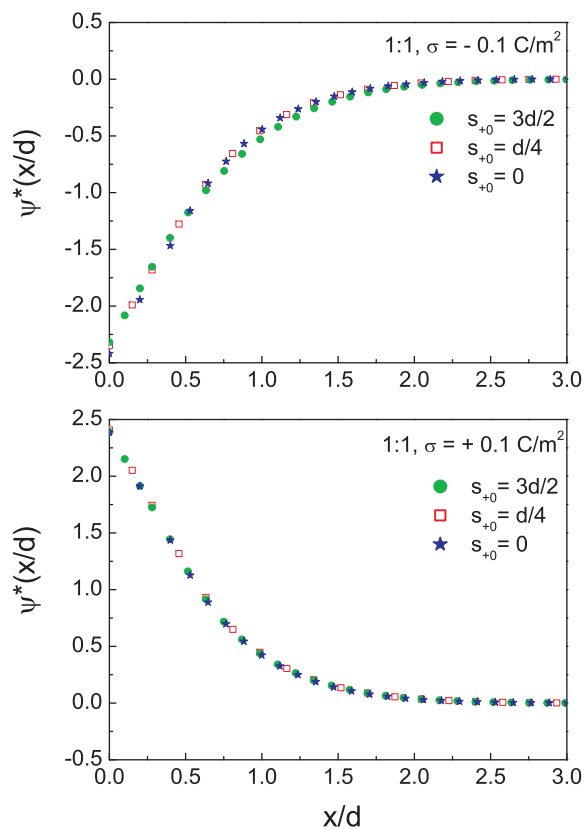}
}
\caption{(Color online) Reduced mean electrostatic potential $\psi ^{*}(x/d)$ for a 1:1 valency electrolyte system
at concentration $c = 1$~mol/dm$^{3}$ with reduced electrode surface charge density $\sigma ^{*} = -0.113$
($\sigma = - 0.1$~C/m$^{2}$) (upper panel), and $\sigma ^{*} = +0.113$
($\sigma = + 0.1$~C/m$^{2}$) (lower panel). The three plots in each panel correspond to
(i) solid circles , $s_{+0}=3d/4$), (ii) open squares, $s_{+0}=d/4)$,
and (iii) stars, $s_{+0}=$ 0. \label{fig4}}
\end{figure}

\begin{figure}[!t]
\centerline{
\includegraphics[width=0.49\textwidth]{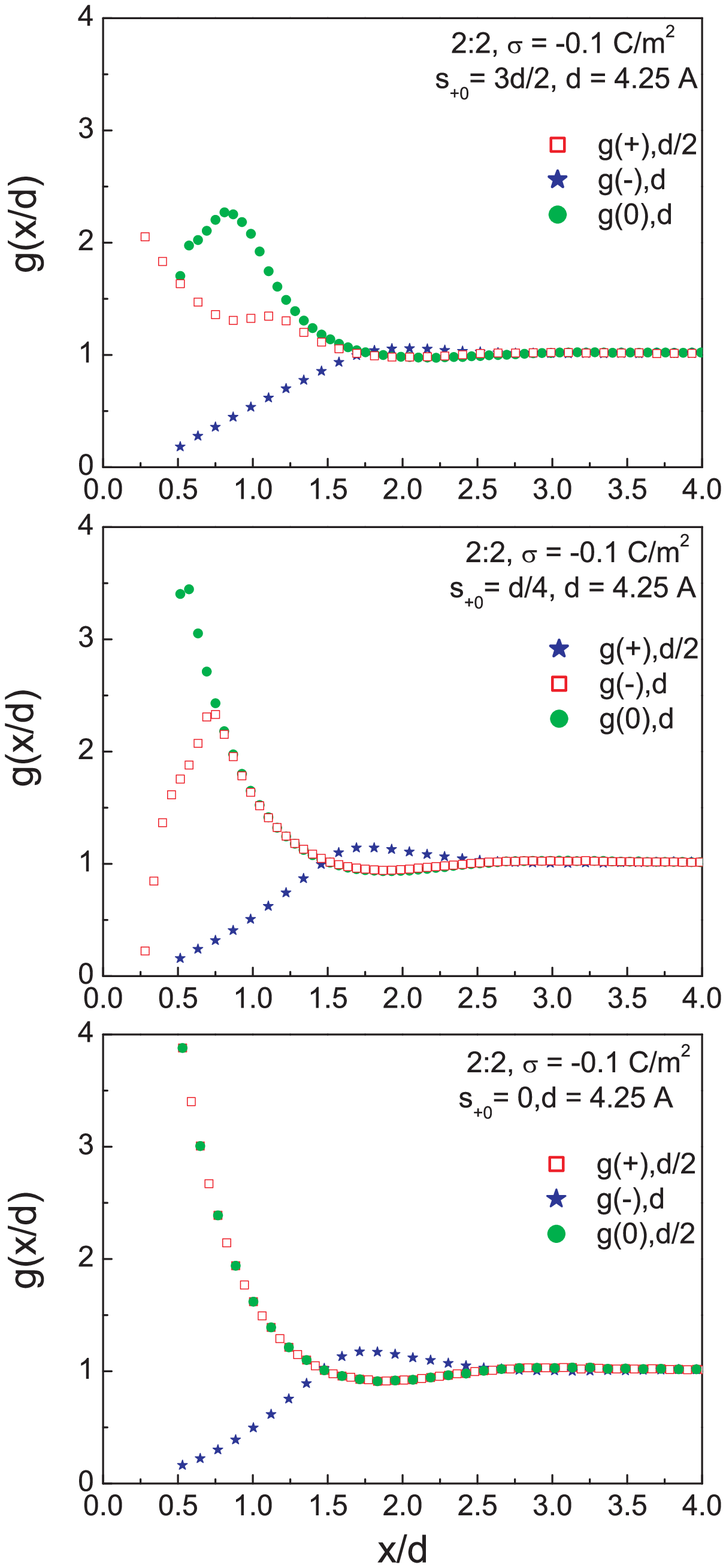}
\includegraphics[width=0.49\textwidth]{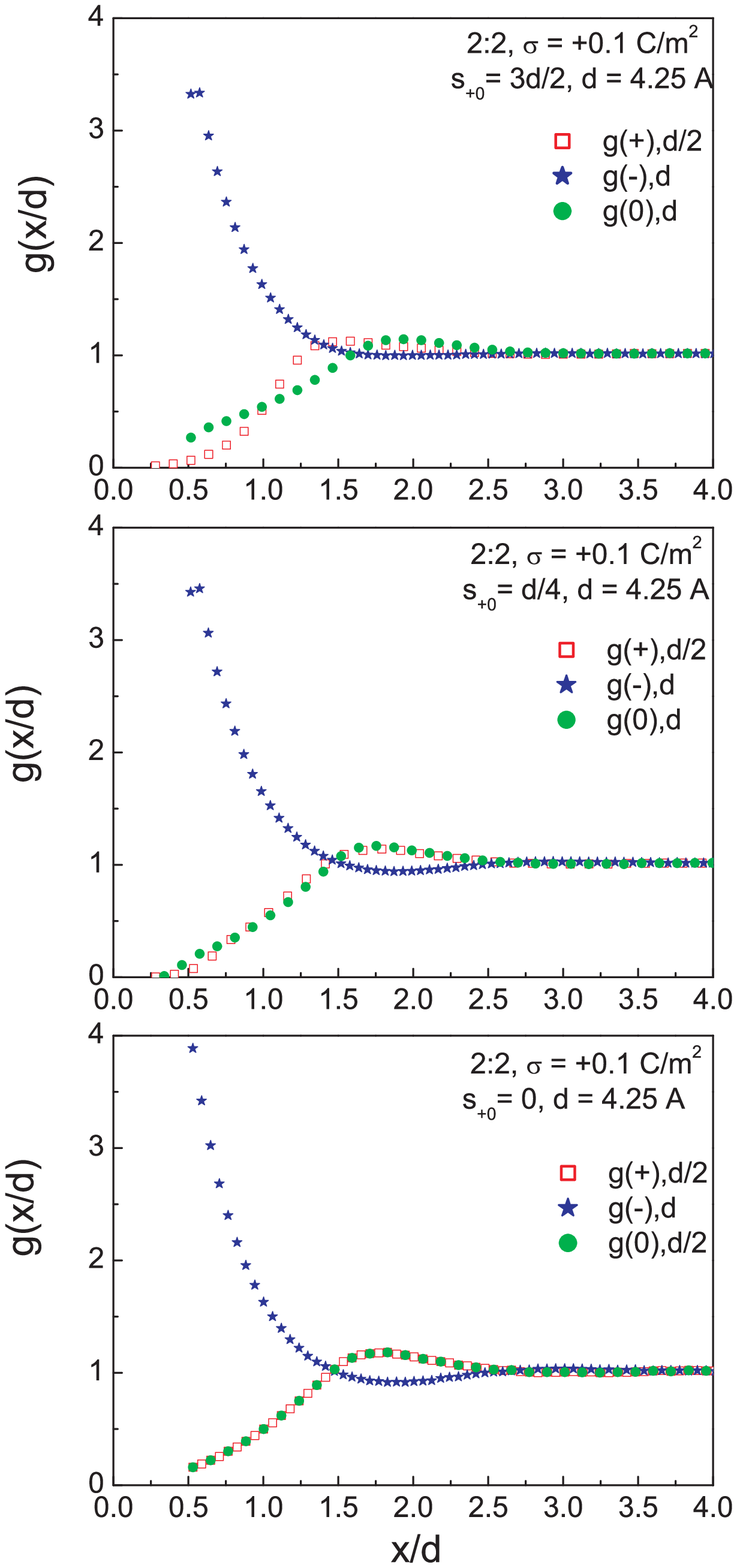}
}
\centering
\parbox[t]{0.49\textwidth}{
\caption{(Color online) Electrode--particle singlet distributions $g_{s}$ for  a 2:2 valency electrolyte system
at concentration $c = 1$~mol/dm$^{3}$ and reduced electrode surface charge density $\sigma ^{*} =-0.113$
($\sigma = - 0.1$~C/m$^{2}$). Rest of the notation and legend as in figure~\ref{fig2}. \label{fig5}}
}
\parbox[t]{0.49\textwidth}{
\caption{(Color online) Electrode--particle singlet distributions $g_{s}$ for  a 2:2 valency electrolyte system
at concentration $c = 1$~mol/dm$^{3}$ and reduced electrode surface charge density $\sigma ^{*} = +0.113$
($\sigma = + 0.1$~C/m$^{2}$). Rest of the notation and legend as in figure~\ref{fig2}. \label{fig6}}}
\end{figure}

\begin{figure}[!t]
\centerline{
\includegraphics[width=0.5\textwidth]{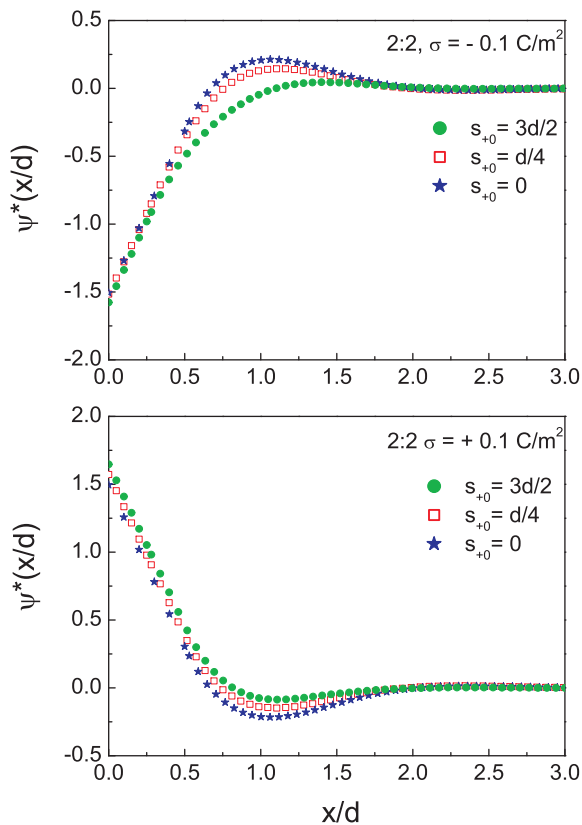}
}
\caption{(Color online) Reduced mean electrostatic potential $\psi ^{*}(x/d)$ for a 2:2 valency electrolyte system
at concentration $c = 1$~mol/dm$^{3}$ with reduced electrode surface charge density $\sigma ^{*} = -0.113$
($\sigma = - 0.1$~C/m$^{2}$) (upper panel), and $\sigma ^{*} = +0.113$
($\sigma = + 0.1$~C/m$^{2}$) (lower panel). Rest of the notation and legend as figure~\ref{fig4}.
\label{fig7}
}
\end{figure}

In figures~\ref{fig2}--\ref{fig4}, results are presented for $Z=1$ for $\sigma=\pm 0.1$~C/m$^2$, which corresponds to $\sigma^*=\pm 0.113$,
whereas the results shown in figures~\ref{fig5}--\ref{fig7} are for these same values of $\sigma$, or $\sigma^*$,
but for $Z=2$.  Our main interest in this work is the case of spheres with an off-center charge, $s_{+0}=d/4$.
These are shown in the middle panels of the density profiles in figures~\ref{fig2}, \ref{fig3}, \ref{fig5}, \ref{fig6}, and in both panels
of the potential profiles in figures~\ref{fig4} and \ref{fig7}.
Results are also given for $s_{+0}=3d/4$ , that is, a dimer consisting of tangentially touching spheres, and
$s_{+0}=0$, which is the well studied RPM. The results for these two cases are not new,
but the fact that these results represent the starting point and the end point in our quest
for an off-center charged ion model, is a valuable internal consistency test of the new code.
Furthermore, comparison of the results for various values
of $s_{+0}$, especially as the degree of fusion increases,
makes the effect of fusing the cation into the neutral sphere physically more apparent.

Considering first figure~\ref{fig2}, the case of a negative surface charge, we notice in general that
being counterions, the cations are attracted to the negatively charged electrode and bring
the neutral spheres with them.  On the other hand, the anions are the coions and are repelled from the electrode.
In the no fusion situation, $s_{0+}=3d/4$ (top panel), the profile of the neutral sphere of the
dimer has a peak for $x$ slightly less that $d$. The charged head of the dimer can approach the
electrode more closely than can the neutral tail of the dimer since the positive sphere has a
smaller diameter than the neutral one and the dimer can rotate. At the other extreme we have
the case of complete fusion such that $s_{+0}=0$  (bottom panel), that is, the RPM case, and the distributions
of the positive and neutral spheres become identical because their centers are
coincident so that the neutral spheres have effectively disappeared. The middle
panel of this figure corresponds to $s_{+0}=d/4$  when the cation
has just disappeared completely into the neutral sphere leading to a RPM with
off-center charged cations, the central theme of the present study. The profiles for the
center of this off-center charged
sphere has a peak for $x$ somewhat greater than $d/2$, whereas the profile of the
charge of this off-center charged sphere has its peak for $x$ nearer $3d/4$. These
profiles also indicate that the distance of the charge of the off-center sphere from
the electrode plane can be less than the corresponding distance of the center of the
sphere itself since the sphere can rotate to bring the charge center closer to the wall.
The profiles of the monomer anions are only slightly affected by the value of $s_{+0}$;
this is because the anion population near the negatively charged electrode
is relatively less. This profile seems to be monotonous for all cases.

The case of a positively charged electrode is considered in figure~\ref{fig3}. The striking
feature of the results here is the qualitative similarity of the various profiles
across the panels. This can be understood from the fact that now it is the
negatively charged monomer spheres that are the counterions and are attracted to the
electrode, while the positively charged spheres and their attached neutral spheres are repelled.
The latter feature, in turn, implies that the internal geometry of the cation has
little influence on the double layer structure in the immediate vicinity of the electrode.
The profiles are thus more akin to that for a PM and having relatively less
features than when the electrode is negatively charged. Analogous observations were made
in some of our previous studies (see for example,
reference \cite{34}). It is interesting that the profile of the counterion has a small peak near the contact.
In the course of these calculations we have also examined the cases with a higher surface
charge density where this peak tended to become less conspicuous so that the peak may well be
related to hard core effects. At any rate, it should have little affect on
the potential profile since the potential profile is a first moment.  The potential profile
is also strongly influenced by the charges further from the electrode. The profiles for the
positive and neutral spheres of the dimers are very nearly indistinguishable and are monotonous.
These particles reside far away from the electrode and seemingly behave as a single entity.
It is noted again that in the bottom panel $s_{+0}=0$  for the RPM case,
the $g_{+}$ and $g_{0}$ are identical as expected.

The 1:1 mean electrostatic potential profiles for the negatively and positively charged walls are
illustrated in figure~\ref{fig4}. In either situation the  $s_{+0}$ dependence is
generally weak. For the monovalent case and hence relatively low coupling, the geometry
of the ions gets masked in taking the weighted average of the $g_{s}$ to get the $\psi ^{*}$.
We further note that the neutral species spheres have no bearing on the potential.
These potential profiles are also monotonous.

  The results for $Z=2$, displayed in figures~\ref{fig5}--\ref{fig7}, show the effect of higher
valency on the structure. The singlet distributions reveal oscillations, not seen
with the 1:1 case, and the double layer is more compact. For a
negative electrode (figure~\ref{fig5}), for the no fusion dimer case (top panel) and the off-center charge
fused sphere (middle panel) case, we note again that the positive spheres can approach the electrode
more closely than the neutral spheres for reasons stated earlier. In figure~\ref{fig6} (positive electrode),
the profiles for positively charged and neutral ends of the dimer (top panel) or the profiles for the center
of the off-center charge sphere and the charge-center of this sphere (middle panel) are very
nearly the same as seen earlier in the 1:1 case. The general shape of the profiles is determined
by the nature of counterions, which are spheres in all cases.  However, rotation of the dimer
coions is clearly evident. The dimer rotates so that the charged end is further from the electrode.
This is still true for the off-center charged sphere but to a lesser extent.
In the bottom panels (RPM case) of these two figures, the $g_{+}$ and $g_{0}$ are identical as before for the
1:1 case.

    The $\psi ^{*}$ profiles depicted in figure~\ref{fig7} are of interest. Unlike the monovalent cases
in figure~\ref{fig5}, the potential profiles for the divalent ions are no longer monotonous. Near the electrode,
there is a broad maximum or a shallow minimum in $\psi ^{*}$  for the negative and positive
surface charges, respectively. This is the phenomenon of \emph{charge inversion} or \emph{overscreening},
which refers to the physical situation where a layer of counterions
predominates near the electrode and over-compensates the charge on the electrode. This is a common
occurrence when divalent ions are present. In the present case, this feature manifests for all~$s_{+0}$.

\section{Summary}

    In this paper we have explored through MC simulations an asymmetric variation of the RPM planar
double layer where asymmetry is imparted to the electrolyte by having an off-center
charge for the cations. This is a logical extension of the familiar RPM.
In the model introduced here, the repulsive hard, non-electrostatic, part of
the inter-ionic interaction is spherical and central, and is represented by a
hard sphere potential, which in many ways is the most difficult part of the
interaction but which has been studied extensively and is now quite  well understood.
However, there is no reason to expect that the center of charge of an ion should be
located at the center of mass of the ion.  This would be especially true for the
ions in an ionic liquid.  One way of deriving a theory is to accept the hard sphere
portion of a charged hard sphere fluid as the starting point and to introduce
the electrostatic contribution as a perturbation.  For example, it is possible
to obtain the mean spherical approximation by means of a `ring sum' of
electrostatic terms.  It may well be possible to formulate an analogous theory
when the electrostatic interaction has a center of charge that differs
from the geometric center, or center of mass, of the ion, especially when the
difference between the center of charge and center of mass is small enough
that it can be regarded as a perturbation.

    Another pertinent question in line with the above concerns the generalization
of the Henderson-Blum-Lebowitz (HBL) contact value theorem for PM double layers [40,41]
to include the present off-center charge RPM double layer. Such an extension will
certainly aid in the theoretical development as has been the case with the HBL sum rule
in the double layer literature. But an expression for the osmotic pressure for a bulk
charged hard sphere fluid with an off-center charge does not yet exist so that again a perturbation
approach would be preferable. We hope to take up the issue in the near future.

    A cation with an off-center charge was conceived from a dimer cation formed by a tangentially
tethered pair of a large hard sphere and a smaller charged hard sphere such that the latter
can fuse entirely into the uncharged sphere. The interesting changes that occur as a result
in the structural pattern of the double layer relative to that of the standard RPM makes the
off-center charge RPM promising for the future.

    Variations of the present off-center charge model can be contemplated. For instance, one
could have off-center charges for both the anion and the cation. Still further asymmetry could
be incorporated by having in addition, asymmetric ionic valencies and/or sizes. Work on such
models is in progress.

\section*{Acknowledgements}
Monika Kaja and Stanis{\l}aw Lamperski acknowledge the financial support from the Faculty of Chemistry,
Adam Micklewicz University in Pozna\'n.

\ukrainianpart

\title{Подвійний електричний шар у моделі твердих сфер \\ із нецентральним зарядом}

\author{В. Сильвестр-Алькантара\refaddr{label1},
Л.Б. Буян\refaddr{label1},
С. Ламперский\refaddr{label3},
M. Кайя\refaddr{label3},
Д. Гендерсон\refaddr{label5}}
\addresses{
\addr{label1} Фізичний факультет і лабораторія теоретичної фізики, Університет Пуерто Ріко, Пуерто Ріко, США
\addr{label3} Факультет фізичної хімії, Університет Адама Міцкевича в Познані,  Познань, Польща
\addr{label5} Факультет хімії та біохімії, Університет Бригама Янга, Прово, США
}

\makeukrtitle

\begin{abstract}

Виконано симуляції профілів густини і потенціалу іонів у плоскому електричному подвійному шарі модельного електроліту або іонної рідини.
Іони електроліту або іонної рідини є зазвичай несферичними; у іонних рідинах катіони є молекулярними іонами. У недалекому минулому цю асиметрію моделювали розглядаючи або сфери, які були асиметричними за розміром і/чи валентністю (примітивна модель), або димерні катіони, які утворені за допомогою тангенційно дотичних сфер. Тут ми розглядаємо сферичні іони, які є асиметричними за розміром та моделюють асиметричну форму за допомогою заряду, який розташований не в центрі катіонних сфер, в той час як заряд аніона поміщено в центрі аніонних сфер.
Різні одночастинкові профілі густини і потенціалу порівнюються  з (і) ситуацією з димерами, коли складові сфери димерного катіона тангенційно прив'язані і з (іі) стандартною примітивною моделлю. Результати показують, що  структура подвійного шару  зазнає значного впливу, особливо коли катіон є контріоном. Будучи цікавою сама по собі, ця модель нецентрального заряду може стати корисною для теорій, які розглядають сферичні моделі та вводять нецентральний заряд як збурення.
\keywords електричний подвійний шар, моделювання, теорія функціоналу густини, сфери з нецентральним зарядом
\end{abstract}

\end{document}